\documentclass[twocolumn,showpacs]{revtex4}
\usepackage{amssymb}
\usepackage{makeidx}
\usepackage{amsmath}
\usepackage{graphicx}
\usepackage{epstopdf}
\usepackage{dcolumn}
\usepackage{bm}

\input{tcilatex}

\begin{document}

\title{Induced nonlinear cross sections of conductive electrons scattering
on the charged impurities in doped graphene }
\author{A.K. Avetissian}
\author{A.G. Ghazaryan}
\email{amarkos@ysu.am}
\author{Kh.V. Sedrakian}
\author{B.R. Avchyan}
\affiliation{Centre of Strong Fields, Yerevan State University, 1 A. Manukian, Yerevan
0025, Armenia }
\date{\today }

\begin{abstract}
Relativistic quantum theory of induced scattering of 2D Dirac particles by
electrostatic field of impurity ion (in the Born approximation) in the doped
graphene at the presence of an external electromagnetic radiation field
(actually terahertz radiation, to exclude the valence electrons excitations
at high Fermi energies) has been developed. It is shown that the strong
coupling of massless quasiparticles in the quantum nanostructures to a
strong electromagnetic radiation field leads to the strongly nonlinear
response of graphene, which opens diverse ways for manipulating the
electronic transport properties of conductive electrons by coherent
radiation fields.
\end{abstract}

\maketitle



\section{Introduction}

The physics of graphene \cite{1,2} with the current advanced technologies
opens large research and applied fields including wide spectrum of
investigations from low-energy condensed matter physics to quantum
electrodynamic (QED) effects \cite{3,4} due to exotic features of
quasiparticles in graphene that behave like massless "relativistic" Dirac
fermions (with the Fermi velocity much less than the light speed in vacuum: $%
\mathrm{v}_{F}=10^{8}\mathrm{cm/s}$) and obey a two-dimensional (2D) Dirac
equation \cite{5,6}. These properties of quasiparticles lead to various
applications of graphene to replicate the fundamental nonlinear QED
processes in much weaker electromagnetic (EM) fields compared to common
materials and vacuum where superintense laser fields of ultrarelativistic
intensities are required for observation of nonlinear phenomena such as
production of electron-positron pairs from the vacuum, nonlinear Compton
scattering, multiphoton stimulated bremsstrahlung (SB) etc, proceeding
actually in the current strong and superstrong laser fields \cite%
{7,7a,7aa,7aaa}.

Among the important processes induced by external radiation fields SB is one
of the first stimulated effects at laser-matter interaction revealed
immediately after the invention of lasers \cite{12a}. The latter is basic
mechanism of energy exchange between the charged particles and plane
monochromatic wave in plasma-like media to provide the energy-momentum
conservation law for real absorption-emission processes. What concerns the
electrons elastic scattering on impurity ions in graphene, there are many
papers with consideration of this basic scattering effect which have been
described mainly within the framework of perturbation theory by
electrostatic potential (see, e.g., \cite%
{BornEllastic,Chen2008,BornEl1,BornEl2,BornEl3,TanAdam2007,Katsnelson}).
Regarding the SB process in graphene at moderate intensities of stimulated
radiation, in case of its linear absorption by electrons (or holes), at the
present time there are extensive investigations carried out in the scope of
the linear theory, see, e.g. \cite{7b,7b1,7b2,7b3,Zhu2014}. The linear
absorbance of single layer graphene from infrared up to visible spectral
range is $\pi \alpha \approx 2.3\%$ which depends solely on the
fine-structure constant $\alpha $\ $=1/137$ \cite{37}. Consequently, taking
into account the graphene thickness, the absorption coefficient will be $%
10^{6}cm^{-1}$, which is a remarkably high value for absorption effect. On
the other hand, high EM radiation absorption by ultrasmall volumes is a very
challenging property for shielding materials used in nanoelectronics,
aviation, and space industry, where strict requirements on lightness and
smallness of materials exist. In this aspect, due to its unique properties
graphene\textit{\ }can act as a THz emitting device and/or shielding
material for nanodevices, enabling to bridge so-called THz gap \cite{18,19}.%
\textit{\ }Meanwhile, consideration of multiphoton SB process in graphene in
moderately strong laser fields as a basic mechanism for radiation absorption
towards the mentioned applications is absent up to now. The latter is given
in the current paper.

In general, the interaction of a free electron with the EM wave is described
by the dimensionless relativistic invariant parameter of intensity $\xi
=eE_{0}\lambda /(2\pi mc^{2})$\ \cite{7}, which represents the wave electric
field (with amplitude $E_{0}$)\ work on a wavelength $\lambda $\ in the
units of electron rest energy. Particularly for THz photons $\hbar \omega
\sim 0.01$\ $\mathrm{eV}$, multiphoton effects take place at $\xi \sim 1$\
that corresponds to intensities $I_{\xi }\sim 10^{14}$\ $\mathrm{Wcm}^{-2}$,
while the massless electron-wave interaction in graphene is characterized by
the dimensionless parameter $\chi =ev_{F}E_{0}/\left( \hbar \omega
^{2}\right) $\ \cite{7a}, which represents the work of the wave electric
field on a period $1/\omega $\ in the units of photon energy $\hbar \omega $%
. Depending on the value of this parameter $\chi $, three regimes of the
wave-particle interaction may be established: $\chi \ll 1$\ -- that
corresponds to one-photon interaction regime \cite{9,10,11}, $\chi \gg 1$ --
which is the static field limit of superpower fields in QED or Schwinger
regime \cite{12}, and $\chi \geqslant 1$\ -- is the multiphoton interaction
regime \cite{7a} with the corresponding intensity $I_{\chi }=\chi ^{2}\times
3.07\times 10^{11}$\ $\mathrm{Wcm}^{-2}\mathrm{[\hbar \omega /eV]}^{4}$.
Comparison of this intensity threshold with the analogous one for the free
electrons or situation in common atoms shows the essential difference
between the values of these thresholds: $I_{\xi }/I_{\chi }\sim 10^{11}$.
Thus, for realization of multiphoton SB in graphene one can expect $10^{11}$%
\ times smaller intensities than for SB in atoms \cite{7aa}, \cite{7aaa}, 
\cite{12a,12aa,13}.

In the present work, the influence of multiphoton effects in SB\ absorption
process with moderately strong laser fields is considered. Here the selected
frequency range of terahertz radiation excludes the valence electrons
excitations at high Fermi energies.

In Sec. I the scattering rates and total multiphoton cross-sections for SB
of conduction electrons in graphene have been obtained. The analytic
formulas in case of screened Coulomb potential have been analyzed
numerically in Sec. II. We have also present results for the angular
dependence of scattered Fermi electrons on the laser radiation intensities.
Conclusions are given in Sec. III.

\section{Multiphoton amplitudes and cross-sections of SB in graphene}

Below we will develop the relativistic scattering theory for the 2D Dirac
fermions on arbitrary electrostatic potential $U(r)$ of an impurity ion in
doped graphene and interacting with an external EM\ wave field of moderate
intensities. To exclude the valence electrons excitations at high Fermi
energies in graphene, we will assume for a EM\ wave actually a terahertz
radiation.

Let us determine the scattering Green function formalism in the Born
approximation by potential $U(r)$. Note that the first nonrelativistic
treatment of SB in the Born approximation has been carried out analytically
in the work \cite{12a}, and then this approach has been extended to the
relativistic domain \cite{12aa}.

Transition amplitude of SB process in the EM wave field from the state with
the canonical momentum $\mathbf{p}_{0}(p_{x},p_{y})$ to the state with
momentum $\mathbf{p}$ $(p_{x},p_{y})$ in graphene plane ($x,y$) can be
written as:

\begin{equation}
C_{\mathbf{pp}_{0}}=-\frac{i}{\hbar }\int \Psi _{\mathbf{p}}^{+}(\mathbf{r}%
,t)U(\mathbf{r})\Psi _{\mathbf{p}_{0}}(\mathbf{r},t)dtd\mathbf{r,}  \label{1}
\end{equation}%
where bispinor function $\Psi ^{+}$ is the complex conjugation of $\Psi $.

The fermion particle wave function $\Psi _{\mathbf{p}}(\mathbf{r},t)$ in the
strong EM wave field may be presented in the form: 
\begin{equation}
\Psi _{\mathbf{p}}(\mathbf{r},t)=\exp \left( \frac{i}{\hbar }\mathbf{pr}%
\right) f_{\mathbf{p}}(t),  \label{2}
\end{equation}%
with the spinor wave function $f_{\mathbf{p}}$ determined as follows:%
\begin{equation}
f_{\mathbf{p}}(t)=\frac{1}{\sqrt{2S}}\left( 
\begin{array}{c}
1 \\ 
e^{i\Theta (\mathbf{p}+\frac{e}{c}\mathbf{A}(t))}%
\end{array}%
\right) e^{-i\Omega (\mathbf{p},t)},  \label{3}
\end{equation}%
where the instantaneous temporal phase $\Omega (\mathbf{p},t)$ is defined
as: $\Omega (\mathbf{p},t)=\frac{\mathrm{v}_{F}}{\hbar }\int \sqrt{\left(
p_{x}+\frac{e}{c}A_{x}\right) ^{2}+p_{y}^{2}}dt$, the function $\Theta (%
\mathbf{p}+\frac{e}{c}\mathbf{A}(t))$ is the angle between the vectors of
particle kinematic momentum in the EM field $\mathbf{p}=\mathbf{p}\left(
p_{x}+\frac{e}{c}A_{x},p_{y}\right) $ and the wave vector potential $\mathbf{%
A}(t)=-c\int_{0}^{t}\mathbf{E}(t^{\prime })dt^{\prime }$ (unit vector $%
\widehat{\mathbf{e}}$ is directed along the axis $OX\mathrm{)}$, and
parameter $S$ is the quantization area -graphene layer surface area. In
terms of these parameters, the graphene linear dispersion law for
quasiparticles energy-momentum $\mathcal{E}(p)$ defined by the
characteristic Fermi velocity $\mathrm{v}_{F}$, reads: $\mathcal{E}=\pm 
\mathrm{v}_{F}\left\vert \mathbf{p}\right\vert $ $=\pm \mathrm{v}_{F}\sqrt{%
p_{x}^{2}+p_{y}^{2}}$, where the upper sign corresponds to electrons and the
lower sign -to holes.

As is known, the state of an electron in the field of a strong EM wave, and
consequently, the cross section of SB essentially depends on the wave
polarization. Hence we will consider the case of certain polarization of the
wave, let a linear.

Let us first study a single electron scattering on charged impurity in
graphene and interacting simultaneously with an EM radiation field $\mathbf{E%
}(t)$ (it is clear that at such small Fermi velocities of scattered
particles the plane monochromatic wave field of frequency $\omega $ will
turn into the uniform periodic electric field of frequency $\omega $: $E(t)=$
$E_{0}\cos \omega t$) let polarized along the $OX$ axis.

For determination of spinor wave function $f_{\mathbf{p}}$ we will use the
results of the paper \cite{Ishikawa} with the obtained formula for
transition amplitude:

\begin{equation}
C_{\mathbf{pp}_{0}}=-\frac{i}{\hbar }\int f_{\mathbf{p}}^{+}(t)U(\mathbf{r}%
)f_{\mathbf{p}_{0}}(t)\exp \left( \frac{i}{\hbar }\left( \mathbf{p-p}%
_{0}\right) \mathbf{r}\right) dtd\mathbf{r,}  \label{4}
\end{equation}%
or represented in the form:%
\begin{equation}
C_{\mathbf{pp}_{0}}=-\frac{i}{\hbar }\int f_{\mathbf{p}}^{+}(t)\widetilde{U}%
\left( \frac{\mathbf{p-p}_{0}}{\hbar }\right) f_{\mathbf{p}_{0}}(t)dt,
\label{5}
\end{equation}%
i.e., the transition amplitude $C_{\mathbf{pp}_{0}}$ (\ref{5}) is depended
by the Fourier transform of the scattering potential: 
\begin{equation}
\widetilde{U}\left( \frac{\mathbf{p-p}_{0}}{\hbar }\right) =\int \exp \left( 
\frac{i}{\hbar }\left( \mathbf{p-p}_{0}\right) \mathbf{r}\right) U(\mathbf{r}%
)d\mathbf{r.}  \label{6}
\end{equation}%
For the impurity potential of the arbitrary form $\widetilde{U}\left[ \frac{%
\mathbf{p-p}_{0}}{\hbar }\right] $ from the relation (\ref{5}) we have:

\begin{equation*}
C_{\mathbf{pp}_{0}}=-\frac{i}{\hbar }\frac{1}{2S}\widetilde{U}\left[ \frac{%
\mathbf{p-p}_{0}}{\hbar }\right]
\end{equation*}%
\begin{equation}
\times \int \left( 1+e^{i\left[ \Theta (\mathbf{p}_{0}+\frac{e}{c}\mathbf{A}%
(\tau ))-\Theta (\mathbf{p}+\frac{e}{c}\mathbf{A}(t))\right] }\right)
\label{7}
\end{equation}%
\begin{equation*}
e^{-\frac{i}{\hbar }\mathrm{v}_{F}\int_{0}^{\tau }\left[ \sqrt{\left( p_{x}+%
\frac{e}{c}A_{x}\right) ^{2}+p_{y}^{2}}-\sqrt{\left( p_{0x}+\frac{e}{c}%
A_{x}\right) ^{2}+p_{0y}^{2}}\right] dt}d\tau .
\end{equation*}%
In accordance to the transition amplitude (\ref{7}), impurity potential can
be expressed in the following form:%
\begin{equation}
C_{\mathbf{pp}_{0}}=\int B(\tau )e^{-\frac{i}{\hbar }\mathrm{v}_{F}\left(
P-P_{0}\right) \tau }d\tau ,  \label{15}
\end{equation}%
where%
\begin{equation*}
B(\tau )=-\frac{i}{\hbar }\frac{1}{2S}\widetilde{U}\left[ \frac{\mathbf{p-p}%
_{0}}{\hbar }\right]
\end{equation*}%
\begin{equation*}
\times \left( 1+e^{i\left[ \Theta (\mathbf{p}_{0}+\frac{e}{c}\mathbf{A}%
(t))-\Theta (\mathbf{p}+\frac{e}{c}\mathbf{A}(t))\right] }\right)
\end{equation*}%
\begin{equation}
\times e^{-\frac{i}{\hbar }\mathrm{v}_{F}\int_{0}^{\tau }\left[ \left( \sqrt{%
\left( p_{x}+\frac{e}{c}A_{x}\right) ^{2}+p_{y}^{2}}-P\right) -\left( \sqrt{%
\left( p_{0x}+\frac{e}{c}A_{x}\right) ^{2}+p_{0y}^{2}}-P_{0}\right) \right]
dt}  \label{17}
\end{equation}%
is the periodic function of time, and the time-depended modules of the
"quasimomentums" $P_{0},P$\ are defined as:%
\begin{equation*}
P_{0}=\frac{\omega }{2\pi }\int_{0}^{\frac{2\pi }{\omega }}\sqrt{\left(
p_{0x}+\frac{e}{c}A_{x}(t)\right) ^{2}+p_{0y}^{2}}dt,
\end{equation*}%
\begin{equation*}
P=\frac{\omega }{2\pi }\int_{0}^{\frac{2\pi }{\omega }}\sqrt{\left( p_{x}+%
\frac{e}{c}A_{x}(t)\right) ^{2}+p_{y}^{2}}dt.
\end{equation*}
Here making a Fourier transformation of the function $B(t)$\ over $t$,\
using the known relations%
\begin{equation}
B(t)=\sum\limits_{n=-\infty }^{\infty }\widetilde{B}_{n}\exp (-in\omega t),
\label{19}
\end{equation}%
\begin{equation}
\widetilde{B}_{n}=\frac{\omega }{2\pi }\int_{0}^{2\pi /\omega }B(t)\exp
(in\omega t)dt,  \label{20}
\end{equation}%
and carrying out the integration over $t$ in the formula (\ref{15}),\ we
obtain:%
\begin{equation}
C_{\mathbf{pp}_{0}}^{(n)}=2\pi \hbar \widetilde{B}_{n}\delta \left( \mathrm{v%
}_{F}P-\mathrm{v}_{F}P_{0}-n\hbar \omega \right) .  \label{21}
\end{equation}

Within the Born approximation, the differential probability $W_{\mathbf{pp}%
_{0}}$ of SB per unit time, from the electron or hole state with
two-dimensional momentum $\mathbf{p}_{0}$ to a state with momentum $\mathbf{p%
}$ in the phase space $Sd\mathbf{P/}(2\pi \hbar )^{2}$ is described by the
formula:

\begin{equation}
W_{\mathbf{pp}_{0}}=\underset{t\rightarrow \infty }{\lim }\frac{1}{t}%
\left\vert C_{\mathbf{pp}_{0}}\right\vert ^{2}PdPd\theta \frac{S}{(2\pi
\hbar )^{2}},  \label{22}
\end{equation}%
where $d\theta $ is the differential scattering angle (linear).

Dividing the differential probability $W_{\mathbf{pp}_{0}}$ (\ref{21}) of SB
by initial flux density $\mathrm{v}_{F}$\ and integrating over $dP$, we
obtain the differential cross-section of SB for quasiparticles in doped
graphene:

\begin{equation}
\frac{d\Lambda }{d\theta }=\underset{n=-n_{m}}{\sum^{\infty }}\frac{d\Lambda
^{\left( n\right) }}{d\theta },  \label{23}
\end{equation}%
where

\begin{equation}
\frac{d\Lambda ^{\left( n\right) }}{d\theta }=\left. \frac{\left\vert 
\mathbf{P}\right\vert }{\mathrm{v}_{F}}\left\vert \widetilde{B}%
_{n}\right\vert ^{2}\right\vert _{\mathrm{v}_{F}P=\mathrm{v}_{F}P_{0}+n\hbar
\omega }  \label{24}
\end{equation}%
is the partial differential cross-section of $n$-photon SB\ with maximum
number of emitted photons $n_{m}$. The total scattering cross-section $%
d\Lambda /d\theta $\ will be obtained making summation over the photon
numbers in the formula for differential partial cross-sections $d\Lambda
^{\left( n\right) }/d\theta $ (\ref{23}). The latter may be represented in
the form:

\begin{equation*}
\frac{d\Lambda ^{\left( n\right) }}{d\theta }=\left\vert \widetilde{U}\left[ 
\frac{\mathbf{p-p}_{0}}{\hbar }\right] \right\vert ^{2}
\end{equation*}%
\begin{equation*}
\times \left\vert \int_{0}^{T}d\left( \frac{t}{T}\right) \left( 1+e^{i\left[
\Theta (\mathbf{p}_{0}+\frac{e}{c}\mathbf{A}(t))-\Theta (\mathbf{p}+\frac{e}{%
c}\mathbf{A}(t))\right] }\right) \exp (in\omega t)\right.
\end{equation*}%
\begin{equation*}
\times \left. e^{-\frac{i}{\hbar }\mathrm{v}_{F}\int_{0}^{t}\left[ \left( 
\sqrt{\left( p_{x}+\frac{e}{c}A_{x}\right) ^{2}+p_{y}^{2}}-P\right) -\left( 
\sqrt{\left( p_{0x}+\frac{e}{c}A_{x}\right) ^{2}+p_{0y}^{2}}-P_{0}\right) %
\right] dt^{\prime }}\right\vert ^{2}
\end{equation*}%
\begin{equation}
\times \delta \left( P-P_{0}-n\frac{\hbar \omega }{\mathrm{v}_{F}}\right) 
\frac{PdP}{4(2\pi \hbar )\mathrm{v}_{F}^{2}\hbar ^{2}}.  \label{exact1}
\end{equation}%
For the comparison with the known results in particular cases of scattering,
let we obtain the partial differential cross-sections of SB in the case of $%
n=\pm 1$. We will produce the expansion in Eq. (\ref{exact1}) into a Taylor
series and keep only the terms of the first order over the electric field.
Then we can obtain an asymptotic formula for the partial differential
cross-sections of SB process in the weak wave field (linear theory):

\begin{equation*}
\frac{d\Lambda ^{\left( \pm 1\right) }}{d\theta }=\frac{\chi ^{2}}{32\pi 
\mathrm{v}_{F}^{2}\hbar ^{3}}\left\vert \widetilde{U}\left[ \frac{\mathbf{p-p%
}_{0}}{\hbar }\right] \right\vert ^{2}
\end{equation*}%
\begin{equation*}
\times \left\vert \left( p_{x}-p_{0x}\right) \left( 1+e^{i\left[ \Theta (%
\mathbf{p}_{0})-\Theta \left( \mathbf{p}\right) \right] }\right) \right.
\end{equation*}

\begin{equation*}
\left. \mp i\frac{\hbar \omega }{\mathrm{v}_{F}}\left( \frac{p_{0y}}{%
p_{0}^{2}}-\frac{p_{y}}{p^{2}}\right) e^{i\left[ \Theta (\mathbf{p}%
_{0})-\Theta \left( \mathbf{p}\right) \right] }\right\vert ^{2}
\end{equation*}

\begin{equation}
\times \delta \left( P-P_{0}\mp \frac{\hbar \omega }{\mathrm{v}_{F}}\right)
PdP.  \label{111}
\end{equation}

Comparing the $n$-photon cross-section $d\Lambda ^{\left( n\right) }$ (\ref%
{exact1}) of SB process with the elastic one, we conclude that formula (\ref%
{exact1}) at $\mathbf{A}(\tau )=0$ ($n=0$) passes to elastic scattering
cross-section $d\Lambda _{elast}$ \cite{BornEl3}, which is\ the analog of
the Mott formula in 2D scattering theory:%
\begin{equation*}
\frac{d\Lambda _{elast}}{d\theta }=\frac{\left\vert \mathbf{p}%
_{0}\right\vert }{8\pi \mathrm{v}_{F}^{2}\hbar ^{3}}\left\vert \widetilde{U}%
\left[ \frac{\mathbf{p-p}_{0}}{\hbar }\right] \right\vert ^{2}
\end{equation*}%
\begin{equation}
\times \left\vert \left( 1+e^{i\left[ \Theta (\mathbf{p}_{0})-\Theta (%
\mathbf{p})\right] }\right) \right\vert ^{2}.  \label{exact3}
\end{equation}%
The phase term $\left( 1+\exp i\left[ \Theta (\mathbf{p}_{0}+\frac{e}{c}%
\mathbf{A}(\tau ))-\Theta (\mathbf{p}+\frac{e}{c}\mathbf{A}(\tau ))\right]
\right) ^{2}$ in Eq. (\ref{exact1}) at $\mathbf{A}(\tau )=0$ is the overlap
factor 
\begin{equation}
\left( 1+e^{-i\theta _{q}}\right) \left( 1+e^{i\theta _{q}}\right) =2\left(
1+\cos \theta _{q}\right) ,  \label{333}
\end{equation}%
where $\theta _{q}=\Theta (\mathbf{p}_{0})-\Theta (\mathbf{p})$. The term (%
\ref{333}) known as a Berry phase term arising from the inherent sublattice
symmetry, which with a graphene fourfold ground state degeneracy arising
from the spin and valley factors, restricts the carriers from backscattering.

\section{Differential cross-sections of SB on the screened Coulomb potential
of impurity ions in graphene}

Now we utilize Eq. (\ref{exact1}) in order to obtain the differential
cross-section in particular case of SB process on a screened Coulomb
potential of impurity ions in graphene \cite{BornEl2}, \cite{Katsnelson}, 
\cite{Adam2007,Ando1982,Saito,Wang2007}. In accordance with \cite{BornEl2},
the Fourier transform $\widetilde{U}\left( \mathbf{q}\right) =\int U(\mathbf{%
r})e^{-i\mathbf{qr}}d\mathbf{r}$ of a charged impurity center potential%
\textbf{\ }can be written as:%
\begin{equation}
\widetilde{U}\left( \mathbf{q}\right) =\frac{2\pi e^{2}}{\widetilde{\kappa }%
q\epsilon \left( q\right) },  \label{29}
\end{equation}%
where $\epsilon \left( q\right) $ ($q=\left\vert \mathbf{q}\right\vert $) is
the 2D finite temperature static dielectric (screening) function in random
phase approximation (RPA) appropriate for graphene \cite{Wang2007}, given by
the formula%
\begin{equation*}
\epsilon \left( q\right) =1+\frac{q_{s}}{q}
\end{equation*}%
\begin{equation}
\times \left\{ 
\begin{array}{c}
1-\frac{\pi q}{8k_{F}},\quad q\leq 2k_{F} \\ 
1-\frac{\sqrt{q^{2}-4k_{F}^{2}}}{2q}-\frac{q\sin ^{-1}2k_{F}/q}{4k_{F}}%
,\quad q>2k_{F}%
\end{array}%
\right. .  \label{26}
\end{equation}%
Here $k_{F}=\varepsilon _{F}/\hbar \mathrm{v}_{F}$ is 2D Fermi wave vector, $%
q_{s}=4e^{2}k_{F}/\left( \hbar \widetilde{\kappa }\mathrm{v}_{F}\right) $ is
the effective graphene 2D Thomas-Fermi wave vector, and $\widetilde{\kappa }$
$\mathbf{=}$ $\kappa \left( 1+\pi r_{s}/2\right) $ is the\ effective
dielectric constant of a substrate. The ratio of the potential to the
kinetic energy in an interacting quantum Coulomb system is measured by the
dimensionless Wigner-Seitz radius $r_{s}=e^{2}/\kappa \hbar \mathrm{v}_{F}$,
where $\kappa $ is the background lattice dielectric constant of the system, 
$e^{2}/\hbar \mathrm{v}_{F}\simeq 2.18$ is \textquotedblleft effective
fine-structure constant\textquotedblright\ in graphene (in the vacuum).
Since we are interested in actual laser pulses for external EM radiation, at
the consideration of numerical results it is convenient to represent the
differential cross-sections of SB on the charged impurities in the form of
dimensionless quantities.

\begin{figure}[tbp]
\includegraphics[width=.51\textwidth]{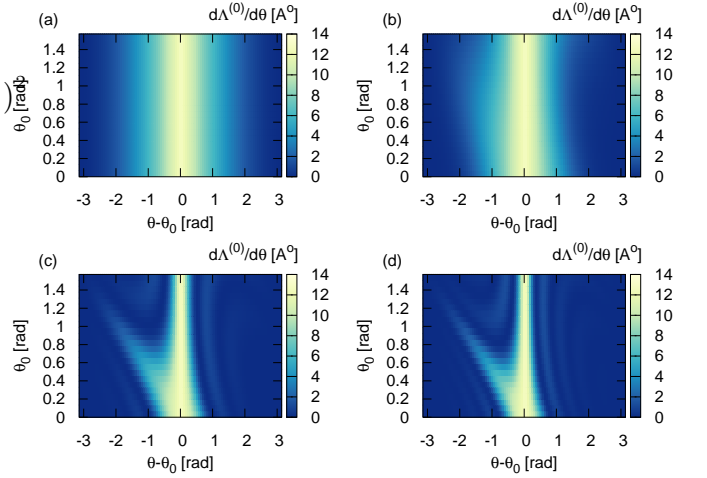}
\caption{{}(Color online) Partial differential cross section $d\Lambda
^{(n)}/d\protect\theta $\ (in angstrom) of SB process for photon number $n=0$
vs the electron deflection angle $\protect\theta -\protect\theta _{0}$\ and
incident angle $\protect\theta _{0}$ for linear polarization of EM wave of
intensities: (a) $\protect\chi _{0}=0$, (b) $\protect\chi _{1}=1$, (c) $%
\protect\chi _{5}=5$, and (d) $\protect\chi _{7}=7$.}
\label{11}
\end{figure}
\begin{figure}[tbp]
\includegraphics[width=.51\textwidth]{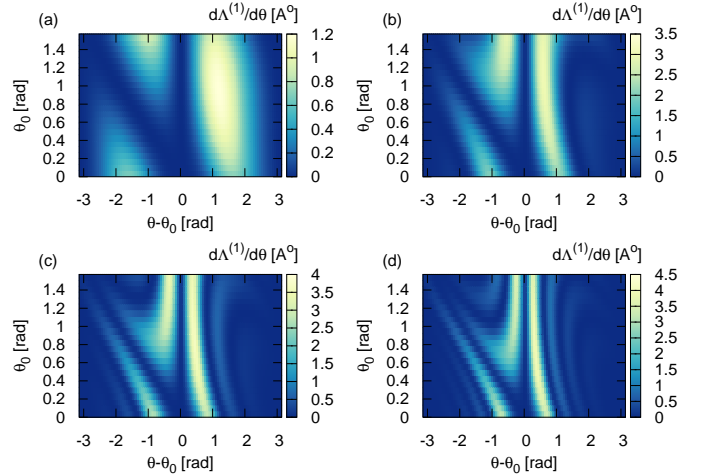}
\caption{ (Color online) Same as Fig. 1 but for photon number $n=1$ and
laser intensities: (a) $\protect\chi _{1}=1$, (b) $\protect\chi _{3}=3$, (c) 
$\protect\chi _{5}=5$, and (d) $\protect\chi _{7}=7$.}
\label{222}
\end{figure}
Taking into account Eqs. (\ref{7}), (\ref{exact1}), and (\ref{29}), we
obtain the following form for the dimensionless partial differential
cross-sections of SB process $d\Lambda ^{\left( n\right) }/d\theta $ in the
field of linearly polarized EM\ wave with the dimensionless vector potential 
$\overline{\mathbf{A}}(t)=-\widehat{\mathbf{e}}\overline{\chi }\sin (2\pi
\tau )$ (unit vector $\widehat{\mathbf{e}}$ is directed along the axis $OX$ $%
\mathrm{):}$

\begin{equation*}
\frac{d\Lambda ^{\left( n\right) }}{\lambda d\theta }=\frac{1}{300}\left( 
\frac{r_{s}}{2+\pi r_{s}}\right) ^{2}\frac{1}{\left\vert \overline{\mathbf{p}%
}\mathbf{-}\overline{\mathbf{p}}_{0}\right\vert ^{2}\epsilon ^{2}\left(
\left\vert \overline{\mathbf{p}}\mathbf{-}\overline{\mathbf{p}}%
_{0}\right\vert \right) }
\end{equation*}%
\begin{equation*}
\times \left\vert \int_{0}^{1}d\tau \left( 1+e^{i\left[ \Theta (\overline{%
\mathbf{p}}-\widehat{\mathbf{e}}\overline{\chi }\sin (2\pi \tau ))-\Theta (%
\overline{\mathbf{p}}_{0}-\widehat{\mathbf{e}}\overline{\chi }\sin (2\pi
\tau ))\right] }\right) \right.
\end{equation*}%
\begin{equation*}
\times \exp \left\{ i2\pi n\tau -2\pi i\int_{0}^{\tau }\left[ \left( \sqrt{%
\left( \overline{p}_{x}-\overline{\chi }\sin (2\pi \tau ^{\prime })\right)
^{2}+\overline{p}_{y}^{2}}-\overline{P}\right) \right. \right.
\end{equation*}%
\begin{equation*}
\left. \left. \left. -\left( \sqrt{\left( \overline{p}_{0x}-\overline{\chi }%
\sin (2\pi \tau ^{\prime })\right) ^{2}+\overline{p}_{0y}^{2}}-\overline{P}%
_{0}\right) \right] d\tau ^{\prime }\right\} \right\vert ^{2}
\end{equation*}%
\begin{equation}
\times \delta \left( \overline{P}-\overline{P}_{0}-n\right) \overline{P}d%
\overline{P}.  \label{exact2}
\end{equation}%
In Eq. (\ref{exact2}) the dimensionless momentum, energy, time, and
relativistic invariant intensity parameter of EM wave introduced as follows:%
\begin{equation*}
\overline{p}_{x,y}=\frac{\mathrm{v}_{F}}{\hbar \omega }p_{x,y},\overline{%
\mathcal{E}}=\frac{\mathcal{E}}{\hbar \omega },
\end{equation*}

\begin{equation*}
\overline{k}_{F}=\frac{\mathcal{E}_{F}}{\hbar \omega },d\tau =\frac{dt}{T},%
\overline{\chi }=\frac{e\mathrm{v}_{F}}{\hbar \omega ^{2}}E_{0}.
\end{equation*}

For numerical analysis of SB\ cross sections in graphene we assume Fermi
energy $\varepsilon _{F}=20\hbar \omega $ ($\varepsilon _{F}\gg n\hbar
\omega $), coherent EM radiation with energy of photons $\hbar \omega =0.01$ 
\textrm{eV }($\lambda =1.24\times 10^{6}\mathrm{\mathring{A}}$), dielectric
environment constant $\kappa =2.5$ for an impurity strength in the presence
of the $\mathrm{SiO}_{2}$ substrate \cite{Wang2007}, Wigner-Seitz radius $%
r_{s}=0.87592$, and effective Fermi temperature $T=0.01\varepsilon _{F}$.

In the Figs. 1-5, the envelopes of partial differential cross sections $%
d\Lambda ^{(n)}/d\theta $ (\ref{exact2}) of SB in graphene for a linearly
polarized EM wave are shown as a function of the angle $\theta _{0}$ between
the vectors of the electron initial momentum and electric field strengths of
a EM wave, $\theta -\theta _{0}$ is the electron deflection angle. These
figures illustrate SB\ at different intensities of stimulated wave. Thus,
Fig. 1 illustrates elastic part of bremsstrahlung ($n=0$), while Figs. 2-5
-- SB for different number of absorbed photons (or emitted photons at $n<0$%
): Fig. 2 -- one-photon SB ($n=1$), Fig. 3 -- two-photon SB ($n=2$), Fig. 4
-- three-photon SB ($n=3$), and Fig. 5 -- four-photon SB ($n=4$),
respectively. The angular dependences of partial differential cross sections
in these figures are displayed for intensities: in Fig. 1 (a) at $\chi =0$,
(b) $\chi =1$ ($I_{\chi }=3\times 10^{3}$ $\mathrm{Wcm}^{-2}$), (c) $\chi =5$
($I_{\chi }=7.7\times 10^{4}$ $\mathrm{Wcm}^{-2}$), and (d) $\chi =7$ ($%
I_{\chi }=1.5\times 10^{5}$ $\mathrm{Wcm}^{-2}$), and in Figs. 2-5: (a) $%
\chi =1$, (b) $\chi =3$ ($I_{\chi }=2.7\times 10^{4}$ $\mathrm{Wcm}^{-2}$),
(c) $\chi =5$, and (d) $\chi =7$, respectively. As one can see, the angular
distribution becomes more asymmetrical with the increasing of the wave
intensity. The maximum values of the cross sections correspond to different
values of the deflection angle $\theta -\theta _{0}$.

In Fig. 6 we plot the dependence of envelopes of partial cross sections $%
n\Lambda ^{(n)}$ for a linearly polarized wave upon the number of emitted or
absorbed photons. The envelopes are obtained via integrating of the partial
differential cross section of SB process $d\Lambda ^{(n)}/d\theta $ (\ref%
{exact2}) over scattering angle of the outgoing electron for diverse laser
intensities. The angle $\theta _{0}$ between the initial electron momentum
and wave electric field is taken to be: $0$, $\pi /6$ $\mathrm{rad}$, and $%
\pi /2$ $\mathrm{rad}$. As it was expected with the increasing of laser
intensity the multiphoton effect becomes dominant compared to the one-photon
scattering in linear theory (\ref{111}). For THz photons, the multiphoton
interaction regime in graphene can be achieved already at the intensities $%
I_{\chi }\sim 10^{3}$ $\mathrm{Wcm}^{-2}$. Thus, for these intensities
multiphoton SB process opens new channels for the wave absorption, and we
can expect strong deviation of absorbance of a single layer doped graphene
from linear one, which for frequencies smaller than Fermi energy is zero 
\cite{37}.

\begin{figure}[tbp]
\includegraphics[width=.51\textwidth]{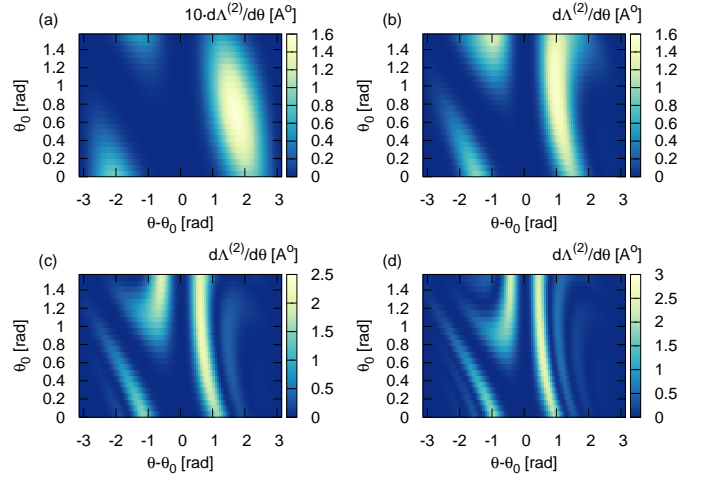}
\caption{{}(Color online) Same as Fig. 2 but for photon number $n=2$.}
\label{33}
\end{figure}
\begin{figure}[tbp]
\includegraphics[width=.51\textwidth]{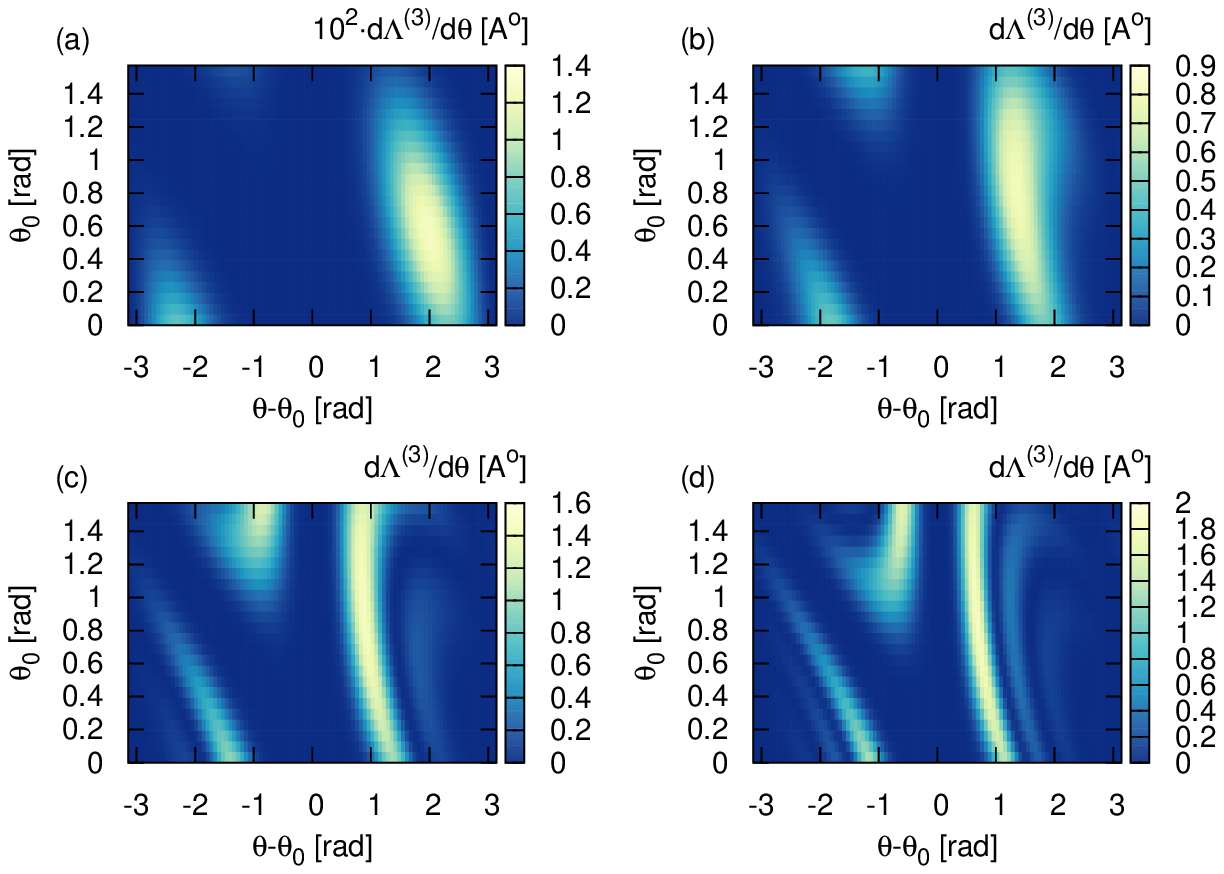}
\caption{{}(Color online) Same as Fig. 2 but for photon number $n=3$.}
\label{44}
\end{figure}
\begin{figure}[tbp]
\includegraphics[width=.51\textwidth]{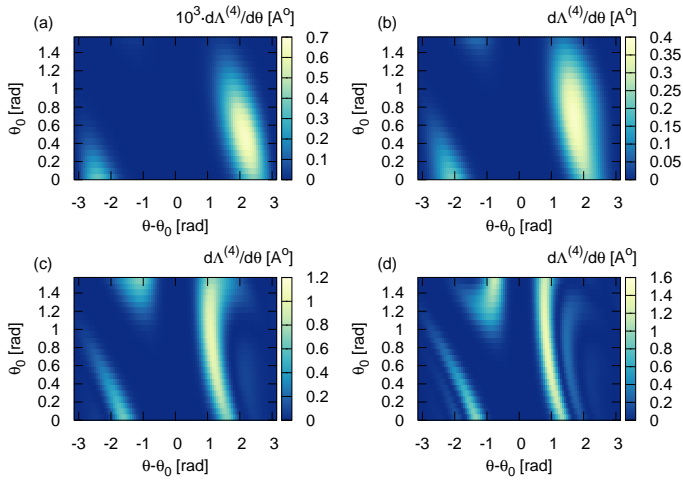}
\caption{{}(Color online) Same as Fig. 2 but for photon number $n=4$.}
\label{55}
\end{figure}
In Fig. 7 we display laser-modified elastic cross section $\Lambda ^{(0)}$
versus intensity parameter for several initial angle $\theta _{0}$ between
the electron momentum and wave electric field. As is seen from this figure,
in the presence of strong radiation field elastic cross section is
essentially modified and decreases with the increase of induced radiation
intensity. The latter opens up possibility for manipulating of electronic
transport properties of the doped graphene by coherent radiation field.

\begin{figure}[tbp]
\includegraphics[width=.6\textwidth]{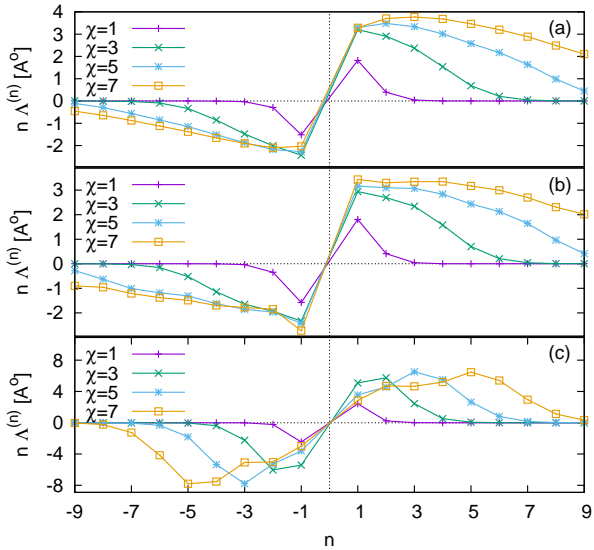}
\caption{{}(Color online) Envelopes of partial absorption-emission cross
sections $n\Lambda ^{(n)}$\ (in angstrom) as a function of the number of
emitted or absorbed photons. The angle $\protect\theta _{0}$ between the
initial electron momentum and wave electric field is taken: (a) $0$, (b) $%
\protect\pi /6$ $\mathrm{rad}$, and (c) $\protect\pi /2$ $\mathrm{rad}$.}
\label{66}
\end{figure}

\begin{figure}[tbp]
\includegraphics[width=.4\textwidth]{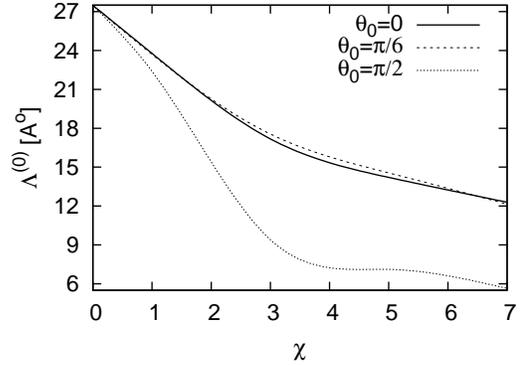}
\caption{ Elastic cross section $\Lambda ^{(0)}$ versus intensity parameter
for several initial angle $\protect\theta _{0}$ between the electron
momentum and wave electric field.}
\end{figure}

\section{Conclusion}

We have presented the theoretical treatment of the multiphoton stimulated
bremsstrahlung process in doped graphene. On the base of the "relativistic"
quantum theory it has been investigated the induced scattering of 2D Dirac
particles on the charged impurity ions of arbitrary electrostatic potential
in the Born approximation and in an external electromagnetic radiation field
(actually terahertz radiation to exclude the valence electrons excitations
at high Fermi energies). The obtained relativistic analytical formulas for
SB at the linear polarization of EM wave have been analyzed numerically for
screened Coulomb potential. The latter shows that SB in graphene in the
presence of strong radiation field is essentially nonlinear, and the
multiphoton absorption/emission processes play significant role already at
moderate laser intensities. For these intensities multiphoton SB process
opens new channels for the wave absorption, which will essentially modify
absorbance of single layer doped graphene compared to the case of linear
absorbance (this problem will be presented in the coming paper). Besides,
the laser-modified elastic cross section is substantially modified and
decreases with the increase of radiation field intensity. The latter opens
up possibility for manipulating of electronic transport properties of the
doped graphene by coherent radiation field.

\begin{acknowledgments}
The authors are deeply grateful to the Prof. Hamlet K. Avetissian and Dr.
G.F. Mkrtchian for permanent discussions during the work on the present
paper, for valuable comments and recommendations. This work was supported by
the State Committee of Science MES RA, in the frame of the research project
No. 15T-1C013.
\end{acknowledgments}


\begin{thebibliography}{99}
\bibitem{1} K. S. Novoselov et al., \textquotedblleft Electric field effect
in atomically thin carbon films\textquotedblright ,\ Science \textbf{306 }%
(5696), 666--669 (2004), http://dx.doi.org/10.1126/science.1102896.

\bibitem{2} A. H. Castro Neto et al., \textquotedblleft The electronic
properties of graphene\textquotedblright ,\ Rev. Mod. Phys. \textbf{81}(1),
109--162 (2009), http://dx.doi.org/10.1103/RevModPhys.81.109.

\bibitem{3} M. I. Katsnelson, K. S. Novoselov, and A. K. Geim,
\textquotedblleft Chiral tunnelling and the Klein paradox in
graphene\textquotedblright , Nature Phys. \textbf{2}, 620--625 (2006),
http://dx.doi.org/10.1038/nphys384.

\bibitem{4} M. I. Katsnelson and K. S. Novoselov, \textquotedblleft
Graphene: new bridge between condensed matter physics and quantum
electrodynamics\textquotedblright , Solid State Commun. \textbf{143}(1--2),
3--13 (2007), http:// dx.doi.org/10.1016/j.ssc.2007.02.043.

\bibitem{5} A. K. Geim, \textquotedblleft Graphene: Status and
prospects,\textquotedblright\ Science \textbf{324}(5934), 1530--1534 (2009),
http://dx.doi.org/10.1126/science.1158877.

\bibitem{6} K. S. Novoselov et al., \textquotedblleft Two-dimensional gas of
massless Dirac fermions in graphene\textquotedblright , Nature \textbf{438},
197--200 (2005), http://dx.doi.org/10.1038/nature04233.

\bibitem{7} H. K. Avetissian, \textquotedblright Relativistic Nonlinear
Electrodynamics\textquotedblright , The QED Vacuum and Matter in
Super-Strong Radiation Fields, Springer, the Netherlands, 2016.

\bibitem{7a} H. K. Avetissian et al., \textquotedblleft Creation of
particle-hole superposition states in graphene at multiphoton resonant
excitation by laser radiation,\textquotedblright\ Phys. Rev. B \textbf{85}%
(11), 115443 (2012), http://dx.doi.org/10.1103/PhysRevB.85.115443.

\bibitem{7aa} H. K. Avetissian, A. G. Ghazaryan, G. F. Mkrtchian,
\textquotedblleft Relativistic theory of inverse-bremsstrahlung absorption
of ultrastrong laser radiation in plasma\textquotedblright , J. Phys. B 
\textbf{46}, 205701 (2013), http://dx.doi.org/10.1088/0953-4075/46/20/205701

\bibitem{7aaa} H. K. Avetissian, A. G. Ghazaryan, H. H. Matevosyan, G. F.
Mkrtchian, \textquotedblleft Microscopic nonlinear relativistic quantum
theory of absorption of powerful x-ray radiation in plasma\textquotedblright
, Phys. Rev. E \textbf{92}, 043103 (2015),
http://dx.doi.org/10.1103/PhysRevE.92.043103.

\bibitem{12a} F. V. Bunkin, A. E. Kazakov, M. V. Fedorov, \textquotedblleft
Interaction of intense optical radiation with free electrons
(nonrelativistic case)\textquotedblright , Sov. Phys.-Usp. \textbf{15}, 416
(1973), http://iopscience.iop.org/0038-5670/15/4/R04; M. H. Mittleman,
\textquotedblleft Introduction to the theory of laser-atom
interactions\textquotedblright , Plenum, New York, 1993.

\bibitem{BornEllastic} T. Ando, \textquotedblleft Screening effect and
impurity scattering in monolayer graphene\textquotedblright , J. Phys. Soc.
Jpn. \textbf{75}, 074716 (2006), http://dx.doi.org/10.1143/JPSJ.75.074716.

\bibitem{Chen2008} J.-H. Chen, C. Jang, S. Adam, M. S. Fuhrer, E. D.
Williams, and D.M. Ishigami, \textquotedblleft Charged-impurity scattering
in graphene\textquotedblright , Nature Physics \textbf{4}, 377 (2008),
http://dx.doi.org/10.1038/nphys935.

\bibitem{BornEl1} K. Nomura and A. H. MacDonald, \textquotedblleft Quantum
Hall ferromagnetism in graphene\textquotedblright , Phys. Rev. Lett. \textbf{%
96}, 256602 (2006), http://dx.doi.org/10.1103/PhysRevLett.96.256602.

\bibitem{BornEl2} E. H. Hwang, S. Adam, and S. Das Sarma, \textquotedblleft
Carrier transport in two-dimensional graphene layers\textquotedblright ,
Phys. Rev. Lett. \textbf{98}, 186806 (2007),
http://dx.doi.org/10.1103/PhysRevLett.98.186806.

\bibitem{BornEl3} D. S. Novikov, \textquotedblleft Elastic scattering theory
and transport in graphene\textquotedblright , Phys. Rev. B \textbf{76},
245435 (2007), http://dx.doi.org/10.1103/PhysRevB.76.245435.

\bibitem{TanAdam2007} Y.-W. Tan, Y. Zhang, K. Bolotin, Y. Zhao, S. Adam,
E.H. Hwang, S. Das Sarma, H. L. Stormer, and P. Kim, \textquotedblleft
Measurement of scattering rate and minimum conductivity in
graphene\textquotedblright , Phys. Rev. Lett. \textbf{99}, 246803 (2007),
http://dx.doi.org/10.1103/PhysRevLett.99.246803.

\bibitem{Katsnelson} M. I. Katsnelson, \textquotedblleft Nonlinear screening
of charge impurities in graphene\textquotedblright , Phys. Rev. B. \textbf{74%
}, 201401 R (2006), http://dx.doi.org/10.1103/PhysRevB.74.201401.

\bibitem{7b} D. P. DiVincenzo and E. J. Mele, \textquotedblleft
Self-consistent effective-mass theory for intralayer screening in graphite
intercalation compounds\textquotedblright , Phys. Rev. B \textbf{29}, 1685
(1984), http://dx.doi.org/10.1103/PhysRevB.29.1685.

\bibitem{7b1} N. H. Shon and T. Ando, \textquotedblleft Quantum transport in
two-dimensional graphite system\textquotedblright , J. Phys. Soc. Jpn. 
\textbf{67}, 2421 (1998), http://dx.doi.org/10.1143/JPSJ.67.2421.

\bibitem{7b2} H. Suzuura and T. Ando, \textquotedblleft Crossover from
symplectic to orthogonal class in a two-dimensional honeycomb
lattice\textquotedblright , Phys. Rev. Lett. \textbf{89}, 266603 (2002),
http://dx.doi.org/10.1103/PhysRevLett.89.266603.

\bibitem{7b3} N. M. R. Peres, F. Guinea, and A. H. Castro Neto,
\textquotedblleft Electronic properties of disordered two-dimensional
carbon\textquotedblright , Phys. Rev. B \textbf{73}, 125411 (2006),
http://dx.doi.org/10.1103/PhysRevB.73.125411.

\bibitem{Zhu2014} S. Sun and J.-L. Zhu, \textquotedblleft Impurity spectra
of graphene under electric and magnetic fields\textquotedblright , Phys.
Rev. B \textbf{89}, 155403 (2014),
http://dx.doi.org/10.1103/PhysRevB.89.155403.

\bibitem{37} R. R. Nair, P. Blake, A. N. Grigorenko, K. S. Novoselov, T. J.
Booth, T. Stauber, N. M. R. Peres, and A. K. Geim, \textquotedblleft Fine
Structure Constant Defines Visual Transparency of Graphene\textquotedblright
, Science \textbf{320}, 1308 (2008),
http://dx.doi.org/10.1126/science.1156965

\bibitem{18} J. Liang et al, \textquotedblleft Electromagnetic interference
shielding of graphene/epoxy composites\textquotedblright , Carbon \textbf{47}%
, 922 (2009), http://dx.doi.org/10.1016/j.carbon.2008.12.038.

\bibitem{19} T. Low and P. Avouris, \textquotedblleft Graphene plasmonics
for terahertz to mid-infrared applications\textquotedblright , ACS Nano 
\textbf{8}, 1086 (2014), http://dx.doi.org/10.1021/nn406627u.

\bibitem{9} E. G. Mishchenko, \textquotedblleft Dynamic conductivity in
graphene beyond linear response\textquotedblright ,\ Phys. Rev. Lett. 
\textbf{103}(24), 246802 (2009),
http://dx.doi.org/10.1103/PhysRevLett.103.246802.

\bibitem{10} P. N. Romanets and F. T. Vasko, \textquotedblleft Rabi
oscillations under ultrafast excitation of graphene\textquotedblright ,\
Phys. Rev. B. \textbf{81}(24), 241411(R) (2010),
http://dx.doi.org/10.1103/PhysRevB.81.241411.

\bibitem{11} B. D\'{o}ra et al., \textquotedblleft Rabi oscillations in
Landau-quantized graphene\textquotedblright , Phys. Rev. Lett. \textbf{102},
036803 (2009), http://dx.doi.org/10.1103/PhysRevLett.102.036803.

\bibitem{12} B. D\'{o}ra and R. Moessner, \textquotedblleft Nonlinear
electric transport in graphene: quantum quench dynamics and the Schwinger
mechanism\textquotedblright , Phys. Rev. B. \textbf{81}(16), 165431 (2010),
http://dx.doi.org/10.1103/PhysRevB.81.165431.

\bibitem{12aa} M. M. Denisov, M. V. Fedorov, \textquotedblleft
Bremsstrahlung effect on relativistic electrons in a strong radiation
field\textquotedblright , Sov. Phys. JETP \textbf{26}, 779 (1968).

\bibitem{13} T. R. Hovhannisyan, A. G. Markossian, G. F. Mkrtchian,
\textquotedblleft On the theory of the relativistic cross-sections for
stimulated bremsstrahlung on an arbitrary electrostatic potential in the
strong electromagnetic field\textquotedblright , Eur. Phys. J. D \textbf{20}%
, 17 (2002), http://dx.doi.org/10.1140/epjd/e2002-00110-7.

\bibitem{Ishikawa} K. L. Ishikawa, \textquotedblleft Nonlinear optical
response of graphene in time domain\textquotedblright ,\ Phys. Rev. B. 
\textbf{82}(20), 201402(R) (2010),
http://dx.doi.org/10.1103/PhysRevB.82.201402.

\bibitem{Adam2007} S. Adam, E. H. Hwang, V. M. Galitski, and S. Das Sarma,
\textquotedblleft A self-consistent theory for graphene
transport\textquotedblright , PNAS \textbf{104}, 18392,
http://dx.doi.org/10.1073/pnas.0704772104.

\bibitem{Ando1982} T. Ando, A.B. Fowler , F. Stern, \textquotedblleft
Electronic properties of two-dimensional systems\textquotedblright , Rev.
Mod. Phys. \textbf{54}, 437 (1982),
http://dx.doi.org/10.1103/RevModPhys.54.437.

\bibitem{Saito} R. Saito, G. Dresselhaus, and M. S. Dresselhaus,
\textquotedblleft Physical properties of carbon nanotubes\textquotedblright
, Imperial College Press, London, UK, 1999.

\bibitem{Wang2007} E. H. Hwang and S. Das Sarma, \textquotedblleft
Dielectric function, screening, and plasmons in two-dimensional
graphene\textquotedblright ,\ Phys. Rev. B. \textbf{75}, 205418 2007,
http://dx.doi.org/10.1103/PhysRevB.75.205418.
\end{thebibliography}
\end{document}